\documentclass[12pt]{article}
\usepackage[dvips]{color}
\usepackage{epsfig}
\usepackage{amsmath}
\usepackage{graphicx}

\begin{document}
\title{\bf Jet-quenching of the rotating heavy meson in a ${\mathcal{N}}$=4
SYM plasma in presence of a constant electric field}
\author{J. Sadeghi $^{a,}$\thanks{Email:
pouriya@ipm.ir}\hspace{1mm}and B. Pourhassan $^{a}$\thanks{Email:
b.pourhassan@umz.ac.ir}\\
$^a$ {\small {\em  Sciences Faculty, Department of Physics, Mazandaran University,}}\\
{\small {\em P .O .Box 47416-95447, Babolsar, Iran}}} \maketitle
\begin{abstract}
In this paper, we consider a rotating heavy quark-antiquark
($q\bar{q}$) pair in a ${\mathcal{N}}$=4 SYM thermal plasma. We
assume that $q\bar{q}$ center of mass moves at the speed $v$ and
furthermore they rotate around the center of mass. We use the
AdS/CFT correspondence and consider the effect of external
electromagnetic field on the motion of the rotating meson. Then we
calculate the jet-quenching parameter
corresponding to the rotating meson in the constant electric field.
\end{abstract}
\section{Introduction}
As we know the Maldacena conjecture [1-3] provides useful
mathematical tools for studying complicated problems of QCD at
strong coupling. In this way one of the important subjects is the
motion of charged particles through the strongly coupled thermal
medium. Already the subject of a quark in the thermal plasma at weak
coupling has been well studied in literature [4-10]. But, QCD at the
strong coupling will be a hard problem, however Maldacena conjecture
make it easy. According to the Maldacena conjecture there is the
relation between type IIB string theory in $AdS_{5}\times S^{5}$
space and $\mathcal{N}$=4 super Yang-Mills (SYM) gauge theory on the
4-dimensional boundary of $AdS_{5}$ space. So, this duality will be
a candidate for the studying strongly coupled plasma. In that case,
instead of a heavy quark in the gauge theory, one may consider dual
picture of the heavy quark which is an open string in AdS space.
Also the dual picture of temperature in the gauge theory is a black
hole (black brane) in $AdS_{5}$ space. One of the important fields
in the QCD, which is also interesting to experiment in the LHC and
RHIC [11], is consideration of the moving heavy quark through the
${\mathcal{N}}$=4 Super Yang-Mills thermal plasma [12-18]. Recently,
the same calculations are done for ${\mathcal{N}}$=2 supergravity
thermal plasma [19]. This subject is important because the solutions
of supergravity theory with $\mathcal{N}$=4 and $\mathcal{N}$=8
supersymmetry may be reduced to the solutions of the $\mathcal{N}$=2
supergravity. In Refs. [19] we found that the problem of drag force
in the $\mathcal{N}$=2 supergravity thermal plasma at zero non -
extremality parameter and finite chemical potential [20] is
corresponding to the $\mathcal{N}$=4 SYM plasma for heavy quark.
Another interesting problem is to consider a $q\bar{q}$ pair which
may be interpreted as a meson. As we know the problem of celebrated
Regge behavior of the hadron spectra has been discussed in
literature [21]. Also the meson spectrum obtained so far and
reasonably describing experiment can be seen in the Ref. [22], where
it is found that the angular momentum plays an important role to
obtain the meson spectrum. So, this give us good
motivation to consider the rotating meson.\\
Already, the energy of the moving $q\bar{q}$ pair through the
${\mathcal{N}}$=4 SYM plasma is studied in both rest frames of
thermal plasma and $q\bar{q}$ pair, which relate to each other by a
Lorentz transformation [23, 24, 25]. Authors in [25] found that the
heavy meson in the plasma feels no drag force. Considered system in
that paper was an ideal case. Actually, the $q\bar{q}$ pair may have
more degrees of freedom such as the rotational motion around the
center of mass and the oscillation along the connection axis. In the
Refs. [26, 27] the description of quark-antiquark system instead
single quark well explained. Also the problem of the spinning open
string (meson) in description of the non-critical string/gauge
duality [28] considered. In that paper the relationship between the
energy and angular
momentum of spinning open string for the Regge trajectory of mesons in a QCD-like theory is studied [29, 30].
It is important to note that, in ${\mathcal{N}}$=4 SYM there is no dynamical quark hence no dynamical mesons,
therefore the rotating mesons are non-dynamical in ${\mathcal{N}}$=4 SYM.\\
Already a rotational motion for the $q\bar{q}$ pair considered, then
the momentum densities for such a system calculated [31]. In that
case their method was different with Refs. [28] and [32]. In the
Ref. [32] authors considered a rotating quark and calculated drag
force on a test quark moving through the plasma. In order to obtain
the drag force one needs to calculate the components of the
energy-momentum density $\Pi_{X}^{1}$ and $\Pi_{Y}^{1}$. We assumed
that $q\bar{q}$ pair moves at a constant speed $v$ along $X$
direction, and also rotates around the center of mass. In [31]
authors obtained the effect of the rotational motion in the
energy-momentum components of the heavy $q\bar{q}$ pair with the
non-relativistic velocity. Therefore, they are determined the motion
of the heavy $q\bar{q}$ pair more exactly. As we know, in the case
of the single quark, the energy of quark goes from D-brane in to the
black hole. In the case of quark and antiquark the currents of
energy and momentum coming from two points on D-brane. These
components cancel each other on the top of the string and therefore
effectively the heavy $q\bar{q}$ pair feels no drag force, so the
string can save its shape in the quark-gluon plasma. In this paper
we would like to consider the effect of constant electromagnetic
field on the motion of rotating heavy meson through the
${\mathcal{N}}$=4 SYM plasma. Also we are going to use results of
the [31] to calculate jet-quenching parameter, which is one of the
interesting properties of the strongly coupled plasma [33-38]. The
jet-quenching parameter supplies a measurement of the dispersion of
the plasma. This quantity commonly obtained by using perturbation
theory, but in here by using the AdS/CFT correspondence we find
jet-quenching parameter in non-perturbative quantum field
theory.\\
This paper organized as the following. In the section 2 we review
some important results of the Ref. [31] for completeness. In the
section 3 we add an external constant electromagnetic field and
obtain the motion of the rotating string. In the section 4 we
calculate the jet-quenching parameter in this system and finally in
the section 5 we summarized our results and give some suggestions
for future works.
\section{Rotating String Equation of Motion}
The dual picture of a meson in the CFT is a string in the AdS space
which both endpoints of it live on the D-brane. This configuration
illustrates the strong interaction between two quarks due to a tube
of gluon field and describes the confinement mechanism in QCD. In
the classical level, these states can be regarded as the rotation of
the system. In another word we will dealing with the spinning open
string.
The spinning string  is interesting because it is dual picture of the rotating meson.
In the Fig.1 we show configuration of the rotating string in the AdS space.\\
\begin{tabular*}{2cm}{cc}
\hspace{0.77cm}\includegraphics[scale=0.5]{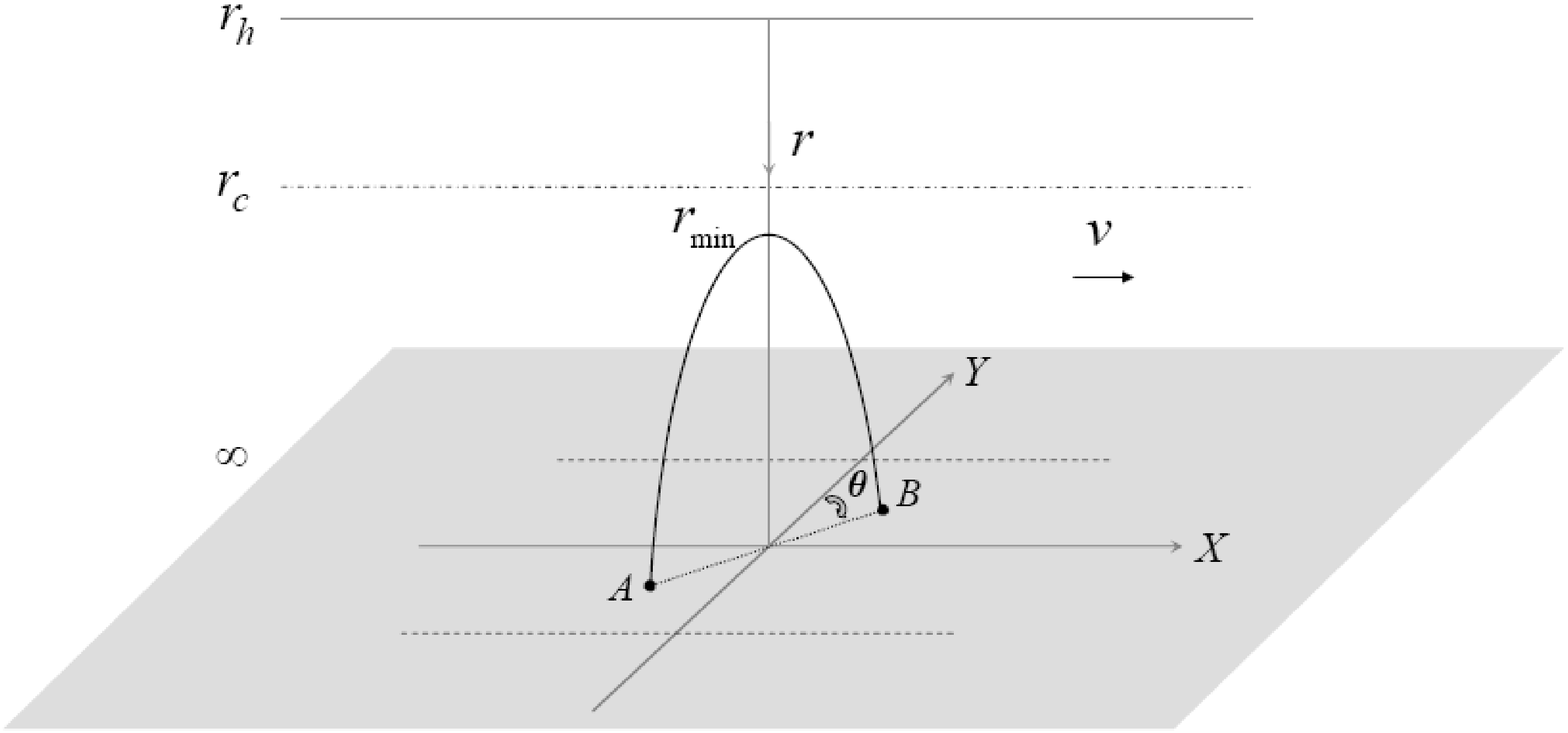}\\
\end{tabular*}\\
Figure 1: A rotating $\cap$ - shape string dual to a $q\bar{q}$ pair
which can be interpreted as a meson. The points $A$ and $B$
represent quark and antiquark with separating length $l$. The radial
coordinate $r$ runs from $r_h$ ( black hole horizon radius) to
$r=r_{m}$ ($r=\infty$)on the $D$-brane. The $r_c$ is the critical
radius, obtained for single quark solution. The $r_{min}$ is turning
point of string and one can obtain $r_{min}\geq r_c$. The parameter
$\theta$ is assumed to be the angle with $Y$ axis and the string
center of mass moves along $X$ axis at velocity
$v$.\\\\
From Maldacena dictionary, we know that, adding temperature to the
system is equal to the existence of a black hole in the center of
$AdS$ space. For the dual picture of ${\mathcal{N}}$=4 SYM plasma
there is the $AdS_{5}$ black hole solution which is given by [25],
\begin{eqnarray}\label{s1}
ds^{2}&=&\frac{1}{\sqrt{H}}(-hdt^{2}+d\vec{x}^{2})+\frac{\sqrt{H}}{h}dr^{2},\nonumber\\
h&=&1-\frac{r_{h}^{4}}{r^{4}},\nonumber\\
H&=&\frac{R^{4}}{r^{4}},
\end{eqnarray}
where  $R$ and $r_{h}$ are curvature radius of AdS space and radius
of black hole horizon respectively,
also the motion direction described by $\vec{x}: (X, Y, Z)$. So we choose motion axis as $X$ and $Y$ (X-Y plan, $Z=0$).\\
We know the dynamics of the open string is described by the
Nambu-Goto action,
\begin{equation}\label{s2}
S=T_{0}\int{dtdr\mathcal{L}},
\end{equation}
where we used static gauge ($\sigma=r$ and $\tau=t$). Therefore the
lagrangian density of system is given by,
\begin{eqnarray}\label{s3}
{\mathcal{L}}&=&-\sqrt{-g}\nonumber\\
&=&-\left[1+\frac{h}{H}({X^{\prime}}^{2}+{Y^{\prime}}^{2})-\frac{1}{h}(\dot{X}^{2}+\dot{Y}^{2})-\frac{1}{H}(\dot{X}^{2}{Y^{\prime}}^{2}+\dot{Y}^{2}{X^{\prime}}^{2}-2\dot{X}X^{\prime}\dot{Y}Y^{\prime})\right]^{\frac{1}{2}},
\end{eqnarray}
where prime and dot denote derivative with respect to $r$ and $t$ respectively,
 also $g\equiv det g_{ab}$, where $g_{ab}$ is the metric on the world
sheet of the string. It is most general lagrangian for the string
which its endpoints lie on $D$-brane. In that case one can obtain
the following expressions for the energy and momentum density
components,
\begin{eqnarray}\label{s4}
\left(\begin{array}{ccc}
\Pi_{X}^{0} & \Pi_{X}^{1}\\
\Pi_{Y}^{0} & \Pi_{Y}^{1}\\
\Pi_{r}^{0} & \Pi_{r}^{1}\\
\Pi_{t}^{0} & \Pi_{t}^{1}\\
\end{array}\right)=-\frac{T_{0}}{H\sqrt{-g}} \left(\begin{array}{ccc}
\dot{Y}Y^{\prime}X^{\prime}-\dot{X}{Y^{\prime}}^{2}-\frac{H}{h}\dot{X} & \dot{X}\dot{Y}Y^{\prime}-X^{\prime}{\dot{Y}}^{2}+hX^{\prime}\\
\dot{X}X^{\prime}Y^{\prime}-\dot{Y}{X^{\prime}}^{2}-\frac{H}{h}\dot{Y} & \dot{X}\dot{Y}X^{\prime}-Y^{\prime}{\dot{X}}^{2}+hY^{\prime}\\
\frac{H}{h}(\dot{X}X^{\prime}+\dot{Y}Y^{\prime}) & -\frac{H}{h}({\dot{X}}^{2}+{\dot{Y}}^{2}-h)\\
h({X^{\prime}}^{2}+{Y^{\prime}}^{2}+\frac{H}{h}) & -h(\dot{X}X^{\prime}+\dot{Y}Y^{\prime})\\
\end{array}\right).
\end{eqnarray}
In order to obtain the total energy, total momentum and angular
momentum of the string we use the following relations respectively,
\begin{eqnarray}\label{s5}
E&=&-\int_{r_{min}}^{r_{0}} dr\Pi_{t}^{0},\nonumber\\
P_{X}&=&\int_{r_{min}}^{r_{0}} dr\Pi_{X}^{0},\nonumber\\
P_{Y}&=&\int_{r_{min}}^{r_{0}} dr\Pi_{Y}^{0},\nonumber\\
J&=&\int_{r_{min}}^{r_{0}} dr(X\Pi_{Y}^{0}-Y\Pi_{X}^{0}).
\end{eqnarray}
It is possible to study Regge trajectory by calculation of
$\frac{E^{2}}{J}$. In that case we should determine the motion of
string, it means that we should find
explicit expressions for $x(r)$ and $y(r)$.\\
Now, we are in the place of applying rotational motion to the
string, so we consider the following anstaz as solution of the
equation of motion,
\begin{eqnarray}\label{s6}
X(r, t)&=& vt+x\sin\omega t,\nonumber\\
Y(r, t)&=& y\cos\omega t,
\end{eqnarray}
with constants linear and rotational motion which are denoted by $v$
and $\omega$ respectively. we choose $\theta(t)=\omega t$ as an
angle with $Y$ axis (see Fig. 1).\\
Here, we consider the special case of small velocities to find
momentum densities. So, we consider a moving heavy $q\bar{q}$ with
non-relativistic speed $v$, which rotate by angle $\theta=\omega t$
around the center of mass, therefore we have $\omega\ll 1$. It is
corresponding to motion of the heavy meson with large spin. Indeed
in the very large angular momentum limit a semiclassical
approximation is reliable. In this case, as mentioned above, the
angular velocity of the string is very small and the separation
value of quark and antiquark is very large. In this limits one can
obtain,
\begin{equation}\label{s7}
\sqrt{-g}\approx\left[1-\frac{v^{2}}{h}+\frac{h}{H}
{x^{\prime}}^{2}\sin^{2}\omega
t+\frac{h-v^{2}}{H}{y^{\prime}}^{2}\cos^{2}\omega
t\right]^{\frac{1}{2}},
\end{equation}
and from equation (4) one finds the following expression of momentum
currents,
\begin{eqnarray}\label{s8}
\left(\begin{array}{ccc}
\Pi_{X}^{1}\\
\Pi_{Y}^{1}\\
\end{array}\right)=-\frac{T_{0}}{H\sqrt{-g}}
\left(\begin{array}{ccc}
hx^{\prime}\sin\omega t\\
(h-v^{2})y^{\prime}\cos\omega t\\
\end{array}\right),
\end{eqnarray}
where $\sqrt{-g}$ is given by the equation (7). By using equations
(7) and (8) one can obtain the following expressions,
\begin{eqnarray}\label{s9}
x^{\prime}\sin\omega t&=&\frac{H(h-v^{2})}{h}\Pi_{X}^{1}\sqrt{\frac{1}{(h-v^{2})(hT_{0}^{2}-H{\Pi_{X}^{1}}^{2})-hH{\Pi_{Y}^{1}}^{2}}},\nonumber\\
y^{\prime}\cos\omega
t&=&H\Pi_{Y}^{1}\sqrt{\frac{1}{(h-v^{2})(hT_{0}^{2}-H{\Pi_{X}^{1}}^{2})-hH{\Pi_{Y}^{1}}^{2}}}.
\end{eqnarray}
Then one can fix momentum densities by using the condition
$\frac{y^{\prime}}{x^{\prime}}=\cot\omega t$ at $r=r_{min}$ [31].
The reality condition for single quark solution yields to the
velocity-dependent critical radius [12, 13, 16, 19],
\begin{equation}\label{s10}
r_{c}=\frac{r_{h}}{(1-v^{2})^{\frac{1}{4}}}.
\end{equation}
By using the reality condition in equation (9) for the
quark-antiquark system one can find special radius, $r_{min}$, where
functions $x$ and $y$ are not imaginary. Then, by using square root
quantity in (9) one can obtain,
\begin{equation}\label{s11}
r_{min}=\left[r_{h}^{4}+\frac{b}{2T_{0}^{2}(1-v^{2})}
\left(1-\sqrt{1-\frac{4T_{0}^{2}(1-v^{2})v^{2}r_{h}^{4}R^{4}{\Pi_{X}^{1}}^{2}}{b^{2}}}\right)\right]^{\frac{1}{4}},
\end{equation}
where we define,
\begin{equation}\label{s12}
b\equiv
R^{4}(1-v^{2}){\Pi_{X}^{1}}^{2}+R^{4}{\Pi_{Y}^{1}}^{2}+T_{0}^{2}v^{2}r_{h}^{4}.
\end{equation}
Indeed, the $r_{min}$ is the turning point of the string. It is easy
to check that $r_{min}\geq r_{c}$. If we consider $\Pi_{X}^{1}=0$
($l=0$) the special case of $r_{min}= r_{c}$ will be satisfied. We
note here $r_{min}= r_{c}$ correspond to the single quark solution
[25]. Also one can see that if $v=0$ then $r_{min}=r_{h}$ which is
expected. The turning point $r_{min}$ is an important parameter to
determine
the jet quenching parameter, so we will use relation (11) in the section 4.\\
In the next section we add an electromagnetic field to the
background and will obtain motion of the rotating meson.
\section{Effect of the constant electromagnetic field}
In the previous section we considered the rotating meson through the
thermal plasma without any external field. Now, we introduce a
constant electromagnetic field in the background. The constant
electromagnetic field assumed along the $X$ and $Y$ directions. The
constant electric field and the constant magnetic field denoted by
$B_{01}$ and $B_{12}$ respectively. This procedure for the single
quark in the ${\mathcal{N}}$=4 SYM theory originally studied in
[39]. In this configuration the lagrangian density (3) changes to,
\begin{eqnarray}\label{s13}
-g&=&1+\frac{h}{H}({X^{\prime}}^{2}+{Y^{\prime}}^{2})-\frac{1}{h}
(\dot{X}^{2}+\dot{Y}^{2})-\frac{1}{H}(\dot{X}^{2}{Y^{\prime}}^{2}
+\dot{Y}^{2}{X^{\prime}}^{2}-2\dot{X}X^{\prime}\dot{Y}Y^{\prime})\nonumber\\
&-&\left(B_{01}X^{\prime}+B_{12}(\dot{X}Y^{\prime}-\dot{Y}X^{\prime})\right)^{2}.
\end{eqnarray}
Therefore the equations of motion read as,
\begin{eqnarray}\label{s14}
&&\frac{\partial}{\partial_{r}}
\left[\frac{1}{\sqrt{-g}}\left(\frac{X^{\prime}\dot{Y}^{2}-\dot{X}\dot{Y}Y^{\prime}-hX^{\prime}}{H}
+(B_{01}-B_{12}\dot{Y})\left(B_{01}X^{\prime}+B_{12}(\dot{X}Y^{\prime}-\dot{Y}X^{\prime})\right)\right)\right]\nonumber\\
&&+\frac{1}{\sqrt{-g}}\frac{\partial}{\partial_{t}}
\left[\frac{\dot{X}}{h}+\frac{\dot{X}{Y^{\prime}}^{2}-X^{\prime}Y^{\prime}\dot{Y}}{H}
+B_{12}Y^{\prime}\left(B_{01}X^{\prime}+B_{12}(\dot{X}Y^{\prime}-\dot{Y}X^{\prime})\right)\right]=0,
\end{eqnarray}
and
\begin{eqnarray}\label{s15}
&&\frac{\partial}{\partial_{r}}
\left[\frac{1}{\sqrt{-g}}\left(\frac{Y^{\prime}\dot{X}^{2}-\dot{X}\dot{Y}X^{\prime}-hY^{\prime}}{H}
+B_{12}\dot{X}\left(B_{01}X^{\prime}+B_{12}(\dot{X}Y^{\prime}-\dot{Y}X^{\prime})\right)\right)\right]\nonumber\\
&&+\frac{1}{\sqrt{-g}}\frac{\partial}{\partial_{t}}
\left[\frac{\dot{Y}}{h}+\frac{\dot{Y}{X^{\prime}}^{2}-X^{\prime}Y^{\prime}\dot{X}}{H}
-B_{12}X^{\prime}\left(B_{01}X^{\prime}+B_{12}(\dot{X}Y^{\prime}-\dot{Y}X^{\prime})\right)\right]=0,
\end{eqnarray}
for $X$ and $Y$ respectively. In this case the string motion
described by the following anstaz,
\begin{eqnarray}\label{s16}
X(r, t)&=& v_{x}t+x\sin\omega t,\nonumber\\
Y(r, t)&=& v_{y}t+y\cos\omega t,
\end{eqnarray}
Therefore the equation (7) extends to the following relation,
\begin{eqnarray}\label{s17}
-g\approx1&-&\frac{v_{x}^{2}+v_{y}^{2}}{h}+\frac{h}{H}
({x^{\prime}}^{2}\sin^{2}\omega t+{y^{\prime}}^{2}\cos^{2}\omega
t)+\frac{1}{H}(x^{\prime}v_{y}\sin\omega t-y^{\prime}v_{x}\cos\omega
t)^{2}\nonumber\\
&-&\left[B_{01}x^{\prime}\sin\omega
t+B_{12}(y^{\prime}v_{x}\cos\omega t-x^{\prime}v_{y}\sin\omega
t)\right]^{2}
\end{eqnarray}
Also the energy and momentum currents in the small velocities
approximation obtained as the following,
\begin{eqnarray}\label{s18}
\Pi_{X}^{0}&=&\frac{v_{x}}{h}+\frac{1}{H}(v_{x}{y^{\prime}}^{2}\cos^{2}\omega
t-v_{y}x^{\prime}y^{\prime}\sin\omega t\cos\omega
t)\nonumber\\
&+&B_{12}y^{\prime}\cos\omega t\left(B_{01}x^{\prime}\sin\omega
t+B_{12}(v_{x}y^{\prime}\cos\omega t-v_{y}x^{\prime}\sin\omega
t)\right)\nonumber\\
\Pi_{Y}^{0}&=&\frac{v_{y}}{h}+\frac{1}{H}(v_{y}{x^{\prime}}^{2}\sin^{2}\omega
t-v_{x}x^{\prime}y^{\prime}\sin\omega t\cos\omega
t)\nonumber\\
&+&B_{12}x^{\prime}\sin\omega t\left(B_{01}x^{\prime}\sin\omega
t+B_{12}(v_{x}y^{\prime}\cos\omega t-v_{y}x^{\prime}\sin\omega
t)\right)\nonumber\\
\Pi_{X}^{1}&=&-\frac{h}{H}x^{\prime}\sin\omega
t+\frac{1}{H}
(x^{\prime}v_{y}^{2}\sin\omega t-v_{x}v_{y}y^{\prime}\cos\omega t)\nonumber\\
&+&(B_{01}-B_{12}v_{y})\left(B_{01}x^{\prime}\sin\omega
t+B_{12}(v_{x}y^{\prime}\cos\omega t-v_{y}x^{\prime}\sin\omega
t)\right)\nonumber\\
\Pi_{Y}^{1}&=&-\frac{h}{H}y^{\prime}\cos\omega t+\frac{1}{H}
(y^{\prime}v_{x}^{2}\cos\omega t-v_{x}v_{y}x^{\prime}\sin\omega t)\nonumber\\
&+&B_{12}v_{x}\left(B_{01}x^{\prime}\sin\omega
t+B_{12}(v_{x}y^{\prime}\cos\omega t-v_{y}x^{\prime}\sin\omega
t)\right),
\end{eqnarray}
up to the factor $\frac{T_{0}}{\sqrt{-g}}$. Now one may consider two
special cases as the following. In the simplest case we can choose
$v_{x}=v$ and $v_{y}=0$ which is corresponding to the existence of
only electric field, ie. $B_{12}=0$. Then one may choose the case of
only magnetic
field which implies $B_{01}=0$.\\
In this paper we consider the case of $B_{12}=0$, then for small
velocities one can obtain,
\begin{eqnarray}\label{s19}
x^{\prime}\sin\omega t&=&\frac{(h-v^{2})}{\sqrt{h(h-HB_{01}^{2})}}
\frac{H\Pi_{X}^{1}}{\sqrt{(h-v^{2})[(h-HB_{01}^{2})T_{0}^{2}-H{\Pi_{X}^{1}}^{2}]-(h-HB_{01}^{2})H{\Pi_{Y}^{1}}^{2}}},\nonumber\\
y^{\prime}\cos\omega t&=&\sqrt{\frac{h-HB_{01}^{2}}{h}}
\frac{H\Pi_{Y}^{1}}{\sqrt{(h-v^{2})[(h-HB_{01}^{2})T_{0}^{2}-H{\Pi_{X}^{1}}^{2}]-(h-HB_{01}^{2})H{\Pi_{Y}^{1}}^{2}}},
\end{eqnarray}
which are extension of the equation (9) to the case of the existing
a constant electric field in the background. In that case, similar
to the Ref. [31], one can find the following expression for the
momentum densities,
\begin{eqnarray}\label{s20}
\Pi_{X}^{1}&=&\frac{(h(r_{min})-H(r_{min})B_{01}^{2})\tan^{2}\omega t}{\sqrt{H(r_{min})(h(r_{min})-v^{2})
+\frac{h(r_{min})-H(r_{min})B_{01}^{2}}{H(r_{min})}\tan^{4}\omega t}},\nonumber\\
\Pi_{Y}^{1}&=&\sqrt{\frac{h(r_{min})-v^{2}}{H(r_{min})+\frac{h(r_{min})-H(r_{min})B_{01}^{2}}{H(r_{min})(h(r_{min})-v^{2})}\tan^{4}\omega
t}}.
\end{eqnarray}
Now by using reality condition one can obtain,
\begin{equation}\label{s21}
r_{min}=\left[r_{h}^{4}+\frac{d}{2T_{0}^{2}(1-v^{2})}
\left(1-\sqrt{1-\frac{4T_{0}^{2}(1-v^{2})R^{4}c}{d^{2}}}\right)\right]^{\frac{1}{4}},
\end{equation}
where,
\begin{eqnarray}\label{s22}
d\equiv
R^{4}(1-v^{2})({\Pi_{X}^{1}}^{2}+T_{0}^{2}B_{01}^{2})+R^{4}{\Pi_{Y}^{1}}^{2}+T_{0}^{2}v^{2}r_{h}^{4},\nonumber\\
c\equiv
v^{2}r_{h}^{4}({\Pi_{X}^{1}}^{2}+T_{0}^{2}B_{01}^{2})+R^{4}B_{01}^{2}{\Pi_{Y}^{1}}^{2}.
\end{eqnarray}
We can see that the effect of the constant electric field is
increasing of the radius $r_{min}$. Again in the case of the
$B_{01}=0$ we recover the relation (11) and in the case of
$B_{01}=0$ and $v=0$ we obtain $r_{min}=r_{h}$ which is expected. In
the next section we will use relations (11) and (21) to obtain the
jet-quenching parameter.
\section{Calculation of the Jet-Quenching Parameter}
In ultra-relativistic heavy-ion collisions at LHC or RHIC,
interactions between the high-momentum Parton and the quark-gluon
plasma are expected to lead to jet energy loss, which is called jet
quenching. It is known that the non-perturbative definition of the
jet-quenching parameter may be obtained in terms of light-like
Wilson loop [33]. In that case we use results of [33-38] to obtain
the jet-quenching parameter in our system. At the first we calculate
the jet-quenching parameter without any external field. Then we add
a constant electric field to the system and obtain the jet-quenching
parameter. In order to find the jet-quenching parameter, first by
introducing the new coordinates $x^{\pm}=\frac{1}{\sqrt{2}}(t\pm
x^{1})$ , we rewrite the line element (1) in the light-cone
coordinates,
\begin{equation}\label{s23}
ds^{2}=\frac{1-h}{2\sqrt{H}}
[(dx^{+})^{2}+(dx^{-})^{2}]-\frac{1+h}{\sqrt{H}}dx^{+}dx^{-}+\frac{1}{\sqrt{H}}[(dx^{2})^{2}+(dx^{3})^{2}]
+\frac{\sqrt{H}}{h}dr^{2}.
\end{equation}
We choose word-sheet coordinates as $r(x^{-},\vec{x})$ and use
static gauge $x^{-}=\tau$ with length $L^{-}$ and $\vec{x}=\sigma$
with length $L$. Because of condition $L^{-}\gg L$ we can assume
that the coordinates $r$ is independent of $\tau$ and one can
consider $r(\sigma)$ as world-sheet coordinates, so there is
boundary condition as $r(\pm\frac{l}{2})=\infty$. Other coordinates
assuming to be constant. Now, definition of the jet-quenching
parameter is given by [33],
\begin{equation}\label{s24}
\hat{q}\equiv8\sqrt{2}\frac{S-S_{0}}{L^{-}L^{2}},
\end{equation}
where $S$ obtained from action (2) by using above definitions, so
one can obtain,
\begin{equation}\label{s25}
S=\sqrt{2}T_{0}L^{-}\int_{r_{min}}^{\infty}{\frac{dr}{r^{\prime}}\sqrt{(1-h)(\frac{1}{H}+\frac{r^{\prime2}}{h})}},
\end{equation}
where $r_{min}$ is given by the equation (11). In order to remove
$r^{\prime}$ in the above expression we use energy conservation law
$\mathcal{H}=\mathcal{L}-\frac{\partial\mathcal{L}}{\partial
r^{\prime}}r^{\prime}=Const.\equiv E$, which yield us to the
following relation,
\begin{equation}\label{s26}
r^{\prime2}=\frac{h}{H}(\frac{r_{h}^{4}}{2E^{2}R^{4}}-1),
\end{equation}
which implies that $E^{2}\leq\frac{r_{h}^{4}}{2R^{4}}$. We are
interested in low energy limit where $E\ll1$. This limit is
corresponding to $L\rightarrow0$ limit which is agree with $L\ll
L^{-}$. The action $S_{0}$ in the equation (24) interpreted as
self-energy of the quark and antiquark which is given by [38],
\begin{equation}\label{s27}
S_{0}=2T_{0}L^{-}\int_{r_{min}}^{\infty}dr\sqrt{-g_{--}g_{rr}},
\end{equation}
where $r_{min}$ is given by the relation (11). Inserting (26) in to
the action (25) and expanding for infinitesimal $E$ yields to the
following relation,
\begin{equation}\label{s28}
S-S_{0}=\frac{\sqrt{2}T_{0}L^{-}E^{2}R^{4}}{r_{h}^{2}}\int_{r_{min}}^{\infty}\frac{dr}{\sqrt{r^{4}-r_{h}^{4}}}.
\end{equation}
On the other hand one can integrate (26) and obtain the following
relation between $L$ and constant $E$,
\begin{equation}\label{s29}
\frac{L}{2}=\frac{\sqrt{2}ER^{4}}{r_{h}^{2}}\int_{r_{min}}^{\infty}\frac{dr}{\sqrt{r^{4}-r_{h}^{4}}}=
(\frac{\sqrt{2}ER^{4}}{r_{min}r_{h}^{2}}){}_{2}F_{1}[\frac{1}{4},\frac{1}{2};\frac{5}{4};
\frac{r_{h}^{4}}{r_{min}^{4}}].
\end{equation}
Inserting equations (28) and (29) in the equation (24) leads us to
the expression for the jet-quenching parameter,
\begin{equation}\label{s30}
\hat{q}=\frac{2T_{0}}{R^{4}}\frac{r_{min}r_{h}^{2}}{{}_{2}F_{1}[\frac{1}{4},\frac{1}{2};\frac{5}{4};
\frac{r_{h}^{4}}{r_{min}^{4}}]}.
\end{equation}
It is easy to compare our result with the previous work. In the case
of $v=\omega=0$ one can obtain $r_{min}=r_{h}$ and therefore
hypergeometric function reduced to the gamma function as
${}_{2}F_{1}[\frac{1}{4},\frac{1}{2};\frac{5}{4};
\frac{r_{h}^{4}}{r_{min}^{4}}]_{r_{min}=r_{h}}=\sqrt{\pi}\Gamma(\frac{5}{4})/\Gamma(\frac{3}{4})$,
then by using relations $r_{h}=\pi R^{2}T$,
$R^{2}=\alpha^{\prime}\sqrt{\lambda}$ and
$T_{0}=\frac{1}{2\pi\alpha^{\prime}}$ we recover the famous relation
of the jet-quenching parameter in ${\mathcal{N}}$=4 SYM theory in
the large $N_{c}$ and large-$\lambda$ limits,
$\hat{q}=\frac{\pi^{2}}{a}\sqrt{\lambda}T^{3}$, where
$a\approx1.311$. However in the case of rotational motion the value of the jet-quenching parameter increases.\\
Now, we consider a two form $B_{01} dt \wedge dx_{1}$. In that case
one can obtain,
\begin{equation}\label{s31}
S-S_{0}=\sqrt{2}T_{0}L^{-}E^{2}R^{4}I,
\end{equation}
where
\begin{equation}\label{s32}
I=\int_{r_{min}}^{\infty}\frac{dr}{\sqrt{({r^{4}-r_{h}^{4}})(r_{h}^{4}-B_{01}R^{2}r^{2})}},
\end{equation}
and the radius $r_{min}$ is given by the equation (21). Also one can
obtain,
\begin{equation}\label{s33}
r^{\prime2}=\frac{h}{H}(\frac{r_{h}^{4}-B_{01}R^{2}r^{2}}{2E^{2}R^{4}}-1).
\end{equation}
Therefore the jet-quenching parameter for the rotating heavy meson
through the ${\mathcal{N}}$=4 SYM thermal plasma in a constant
electric field obtained as the following relation,
\begin{equation}\label{s34}
\hat{q}=\frac{2T_{0}}{R^{4}}I^{-1}.
\end{equation}
In order to obtain the explicit expression of the jet-quenching
parameter, including the electric field, we assume that the electric
field is infinitesimal parameter. For the infinitesimal electric
field one can obtain,
\begin{equation}\label{s35}
\hat{q}=\frac{\pi}{\alpha^{\prime}}T^{2}\frac{r_{min}}{{}_{2}F_{1}[\frac{1}{4},\frac{1}{2};\frac{5}{4};
\frac{r_{h}^{4}}{r_{min}^{4}}]}\left(1-\frac{r_{min}}{2\pi^{2}\alpha^{\prime}\sqrt{\lambda}T^{2}}\frac{B_{01}}{{}_{2}F_{1}[\frac{1}{4},\frac{1}{2};\frac{5}{4};
\frac{r_{h}^{4}}{r_{min}^{4}}]}\int_{r_{min}}^{\infty}\frac{r^{2}dr}{\sqrt{r^{4}-r_{h}^{4}}}\right).
\end{equation}
It means that the effect of the constant electric field is
decreasing of the jet-quenching parameter. This is agree with the
fact that electric field decreases the drag force [19]. Therefore we
calculated the jet-quenching parameter in rotating meson system
under effect of the constant electric field. It is easy to check
that the above results are coincide with the previous work, without
electric field, if we cancel the electric field, ie. $B_{01}=0$.
\section{Conclusion}
In this paper we generalized the problem of the rotation meson in
the ${\mathcal{N}}$=4 SYM thermal plasma to the case of the existing
constant electromagnetic field. We considered the constant
electromagnetic field in the background and obtained the lagrangian
density for the small velocities. Then we calculated the effect of
the constant electric field on the motion of the rotating meson and
obtained the momentum densities. We have shown that in the presence
of a constant electric field the distance of the turning point from
the D-brane is smaller than the case of without the electric field.
Then, without presence of any external field we obtained the
jet-quenching parameter for the rotating mesons in terms of the
hypergeometric function and have shown that our results for the
$\omega=0$ are coincide with the case of non-rotating meson. We
found that the rotating mesons have larger jet-quenching parameter
than non-rotating mesons. Finally we calculated the jet-quenching
parameter under effect of a constant electric field.
We found that the electric field decreases the value of the jet-quenching parameter.\\
Here, there are some interesting problem for future works. For
example one can obtain jet quenching parameter [33-38] for rotating
$q\bar{q}$ pair or shear viscosity [40] in other backgrounds such as
${\mathcal{N}}$=2 supergravity [19]. Also it is interesting to
consider the effect of higher derivative terms as in the previous
cases [41-49]. As a recent work [50] one may consider more quarks,
such as four quarks in the baryon through ${\mathcal{N}}$=4 SYM
thermal plasma. At the end, it may be interesting to consider
fluctuations of the quark-antiquark pair and obtain the exact
solution of such a
system.\\\\\\
{\bf Acknowledgment:} We would like to thanks Dr. K. Bitaghsir
Fadafan for useful discussion about the jet-quenching parameter.\\

\end{document}